# Deterministic and Scalable Quantum Light Generation in DNA Origami-Programmed Organic Molecule–MoS₂ Monolayer Hybrids

*Shen Zhao[1,2], Zhijie Li[1,3], Elisabeth Erber[1], Anna Altunina[4], Christoph Sikeler[1], Kenji Watanabe[5], Takashi Taniguchi[6], Anvar S. Baimuratov[1,2,4]\*, Tim Liedl[1]\*, Alexander Högele[1,2], and Irina V. Martynenko[1,4]\**

[1]Faculty of Physics and Center for NanoScience (CeNS),

Ludwig-Maximilians-Universität München

Geschwister-Scholl-Platz 1, 80539 Munich, Germany,

[2]Munich Center for Quantum Science and Technology (MCQST),

Schellingstraße 4, 80799 München, Germany

[3]Institute of Physics, Carl von Ossietzky University, 26129

Oldenburg, Germany

[4]Skolkovo Institute of Science and Technology, Moscow 121205, Russia

[5]Research Center for Electronic and Optical Materials, National Institute for Materials Science, 1-1 Namiki, Tsukuba 305-0044, Japan

[6]Research Center for Materials Nanoarchitectonics, National Institute for Materials Science, 1-1 Namiki, Tsukuba 305-0044, Japan

Corresponding authors:

Anvar S. Baimuratov anvar.baimuratov@physik.uni-muenchen.de
Tim Liedl tim.liedl@physik.lmu.de
Irina V. Martynenko irina.martynenko@physik.uni-muenchen.de

The functionalization of atomically-thin transition metal dichalcogenides (TMDs) with organic molecules is a promising approach for realizing nanoscale optoelectronic devices with tailored functionalities, such as quantum light generation or *p-n* junctions. However, achieving precise control over the molecules' positioning on the 2D material remains a significant challenge. Here, we overcome the limitations of solution- and vapor-deposition methods and use a DNA origami placement technique to spatially arrange thiol molecules on a chip surface at the single-molecule level with high assembly yields. We successfully integrated MoS₂ monolayers with micron-scale thiol–origami patterns, achieving single-photon emission from thiol-induced localized excitons in MoS₂. Our work lays a foundation for the chemical control of quantum emitters in atomically-thin semiconductors and enables the design and production of ultracompact 2D devices for quantum technologies.

## Introduction

Solid-state single-photon emitters (SPEs) provide a scalable and robust pathway for achieving quantum memory in quantum networking applications.[1, 2] Among these, SPEs based on atomically-thin TMDs[3, 4] offer several advantages: their emission wavelengths fall within the visible/near-infrared range, their two-dimensional nature facilitates integration into nanoscale devices, and they exhibit high brightness and narrow linewidths.[5] SPEs in TMDs can be created through defects induced by strain[6, 7, 8, 9, 10] or point defects.[10, 11] These defects are typically fabricated via local bending,[6, 7, 8, 12] mechanical scretching,[9, 13] or ion- or electron beam irradiation.[10, 11, 14] However, a common limitation of these techniques is the difficulty in deterministically creating SPEs at specific locations with reproducible properties. Therefore, developing scalable, deterministic, and precise methods for SPE fabrication remains a significant challenge.

Molecular surface functionalization presents an alternative approach to engineering quantum emitters in low-dimensional materials. For instance, functionalization with organic molecules has been used to introduce quantum defects[15] and achieve single-photon emission[16] in carbon nanotubes. For atomically-thin materials as graphene and TMDs, functionalization with organic molecule has also inspired the design of novel device architectures, including high-performance photodetectors,[17, 18] *p-n* junctions,[19, 20, 21] single-molecule sensors,[22, 23] and catalytic systems.[24] In these devices, organic molecules deposited on the surface of 2D material introduce charge doping[25] or energy-transfer states[26], leading to changes in their optical,[27] electronic, thermoelectric,[28] and magnetic properties.[29] However, assembling such hybrid devices at the molecular scale involves several challenging requirements, including nanometer-precise control of molecule positioning, scalability, and versatility in the choice of molecule. Current fabrication methods offer limited control over the homogeneity and positioning of organic molecules, as well as the stability of the final hybrid structure, which hinders progress in this field. Chemical surface functionalization via the non-covalent[30] or covalent[31] bonding of adsorbed molecules are commonly used to fabricate hybrid structures. However, the solution processing methods such as spin-coating, drop-casting, or vapor-based methods lack full control over molecule concentration and position on a 2D material surface. Conventional photolithography processes including photochemical reactions, chemical etching or laser writing[32, 33, 34] can only pattern layers of organic molecules on top of 2D material.

Structural DNA nanotechnology, particularly DNA origami self-assembly,[35, 36] enables bottom-up fabrication of highly defined, complex two- and three-dimensional nanostructures with single-nanometer resolution.[37, 38, 39, 40, 41, 42, 43] DNA origami is widely used as a molecular scaffold, enabling sub-nanometer precision in positioning molecules and nanoparticles.[44, 45, 46, 47, 48, 49] Kershner et al. developed a DNA origami placement (DOP) technique by combining DNA origami self-assembly with lithographic nanopatterning, allowing for site- and shape-selective deposition of individual DNA origami objects into arrays on patterned substrates.[50] DOP overcomes the limitations of top-down lithography and gives access to high-yield placement and arrangement of individual nanoscale components such as metallic nanoparticles[51, 52, 53] and organic dyes[54, 55] into precise arrays and patterns. Importantly, DNA origami placement yields of more than 90%[53, 55] surpass the single-molecule binding efficiency of 37% imposed by Poisson statistics on traditional single-molecule deposition methods.[56, 57] Progressing from these achievements, an approach that allows the fabrication of molecular patterns for chemical control of quantum emitters in TMDs is highly desirable.

Here, we demonstrate programmable 2D material–organic molecule hybrids with high efficiency and spatial precision. By dry-stamp transferring micron-scale, chemical-vapour-deposited MoS$_2$ monolayers onto the chips patterned with DNA origami triangles bearing thiol molecules, we rationally design hybrid systems with localized exciton states. Cryogenic PL of thiol–origami-MoS$_2$ exhibit $g^{(2)}(0)$ values below 0.5, confirming single-photon emission from thiol-induced localized states in MoS$_2$. The resulting SPEs display nanosecond lifetimes and high stability. We achieve a ~90% yield in single-photon emitter placement. Our method provides a platform for precisely engineering the electronic properties of 2D materials at the nanoscale and opens a path toward producing miniaturized hybrid inorganic–organic devices with enhanced performance.

## Fabrication design

Previous investigations have highlighted the potential for functionalizing MoS$_2$ with thiol-terminated molecules.[58, 59, 60] Here, we precisely position thiol molecules on lithographically patterned substrates using DNA origami triangles with 127-nm-long outer edges as "molecular adaptors" (Fig. 1a). A MoS$_2$ monolayer is subsequently transferred on top of the air-dried thiol–origami pattern, resulting in thiol molecules binding to the MoS$_2$ (Fig. 1b).

We used a variant of the "Rothemund triangle" as it has been successfully positioned on lithographically patterned substrates before.[61] The DNA triangle carries 18 3'-adenine "anchor" strands, each 20 nt long, extending from the triangle every 11 nanometers along both its inner edges for functionalization with 19 nt thymine "linker" strands bearing on their 5' ends a single thiol group (the design of the anchor strands is presented in Supplementary Note 1 and Supplementary Fig. 1).

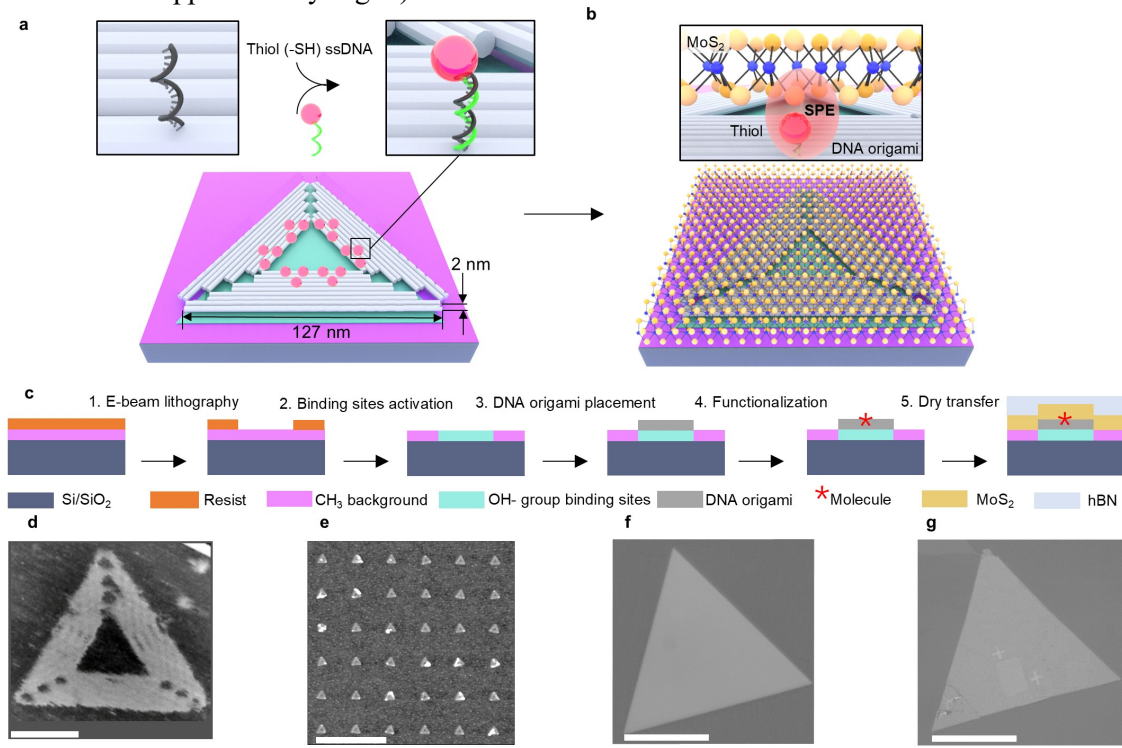

**Figure 1. DNA origami-programmable thiol-MoS$_2$ hybrid assembly. (a)** Schematic illustration of a DNA origami triangle positioned on a hydrophilic, hydroxyl group covered binding site (green) on a hydrophobic, methyl group covered surface of Si/SiO$_2$ chip. DNA origami triangle bears 18 adenine "anchor" strands each 20 nt long for on-surface annealing with 19 nt thymine strands bearing a single thiol group. **(b)** Schematic illustration of a thiol–origami-MoS$_2$ assembly. **(c)** The various steps of the fabrication process of a thiol–origami-MoS$_2$ assembly. **(d)** Liquid mode AFM image of DNA origami triangle in the folding buffer. **(e)** AFM image of a dried Si/SiO$_2$ chip with the DNA triangles positioned onto ~ 120 nm triangular binding sites in a 20 μm-square array with a 250 nm period, fabricated using electron beam lithography. **(f)** Bright-field microscopy

image of the MoS₂ flake on adhesive polycarbonate stamp before transferring onto an origami pattern. (**g**) Bright-field microscopy image of the MoS₂ flake transferred onto the origami pattern. The scale bar in panel (**d**) is 50 nm, in panel (**e**) is 500 nm, and in panels (**f**) and (**g**) is 30 µm.

The various steps of the fabrication process are illustrated in Fig. 1c. First, DNA triangles, designed *in silico*[62, 63] and folded in buffer containing $MgCl_2$ (Fig. 1d), were deposited onto specific binding sites that were created *via* electron-beam (e-beam) patterning of negatively charged hydroxyl groups (green) within a background of hydrophobic methyl groups (pink) on a Si/SO₂ substrate. Negatively charged origami binds strongly to the charged spots via positively charged $Mg^{2+}$ ions from buffer solution while the hydrophobic, thus passivated area, binds origami poorly.

Substrate preparation and DNA triangle placement were performed according to previously established protocols.[53, 61] In brief, Si/SiO₂ chips with square arrays of triangular binding sites were incubated with DNA triangles with build-in anchor strands on the patterned chips at 25°C for 1h in a Tris buffer with 35 mM $MgCl_2$. Surface-bound DNA triangles were functionalized with thiol molecules by incubating the chips with 19-nt thymine strands bearing a single thiol group. After placement and functionalization, the Si/SiO₂ chips with the thiol–origami patterns were air-dried. Atomic force microscopy (AFM) revealed a typical height of 2 nm for the dried DNA triangles, consistent with the height of a DNA duplex (Fig. 1e). In this study, we used square arrays of triangular binding sites with various periods ranging from 170 nm to 1000 nm. Detailed information on the thiol–origami pattern designs, along with AFM characterization and height profiles of the air-dried patterns, is provided in Supplementary Note 2 and Supplementary Figs. 2–3. Consistent with our prior findings,[53] the yield of single origami binding exceeded 90% for all patterns, independent of the array period.

Finally, triangular single-crystal monolayers of MoS₂ covered by hexagonal boron nitride (hBN) flakes (Fig. 1f) were transferred onto a thiol–origami pattern using an adhesive polycarbonate stamp, followed by annealing under vacuum. Fig. 1g depicts a representative optical microscope image of the resulting hybrid assemblies, showing the origami pattern fully covered with MoS₂.

### Optical properties of MoS₂ on thiol–origami pattern

We first aimed to confirm binding between MoS₂ and the origami-bound thiols. To that end, we fabricated dense, periodic square arrays of thiol-functionalized DNA triangles with a 170 nm period, resulting in approximately 40 nm spacing between adjacent DNA triangles (see insets in Fig. 2a and 2b). The thiol–origami pattern design and its AFM characterization are presented in Supplementary Fig. 2. We then transferred a micron-scale, atomically thin MoS₂ flake onto this pattern (Fig. 2a). A control sample without thiol functionalization was also fabricated to rule out interactions between DNA origami and MoS₂.

Fig. 2b depicts room-temperature (RT) photoluminescence (PL) maps of MoS₂ transferred onto the thiol–origami pattern. Both the non-patterned MoS₂ and the MoS₂ deposited on non-thiolated origami patterns exhibited similar PL spectra, characteristic of MoS₂ monolayers on dielectric substrates (Fig. 2c). Specifically, the PL spectra contain a single peak at 1.88 eV with a full-width at half-maximum (FWHM) of 49 meV, corresponding to the lowest excitonic resonance, known as the A exciton.[64] This exciton corresponds to the momentum-direct transition at the K-valleys in the Brillouin zone of the MoS₂ monolayer. Thiol–origami patterning of MoS₂ introduced a new PL peak at 1.83 eV (Fig. 2b, 2c). This new peak is red-shifted by 50 meV compared to the A exciton transition and exhibits half the intensity and a 29% broader FWHM of 68 meV. Notably, this red-shifted peak cannot be attributed to trion states, as the monolayer MoS$_2$ trion binding energy of 34 meV[65, 66] is less than the observed 50 meV energy splitting between the A exciton and the red-shifted peak. This splitting is more consistent with localized exciton states, such as those created by helium ion irradiation.[11] We hypothesize that thiol binding to MoS$_2$ leads to the formation of localized states, potentially serving as the origin of single-photon emission.[67, 68, 69, 70, 71]

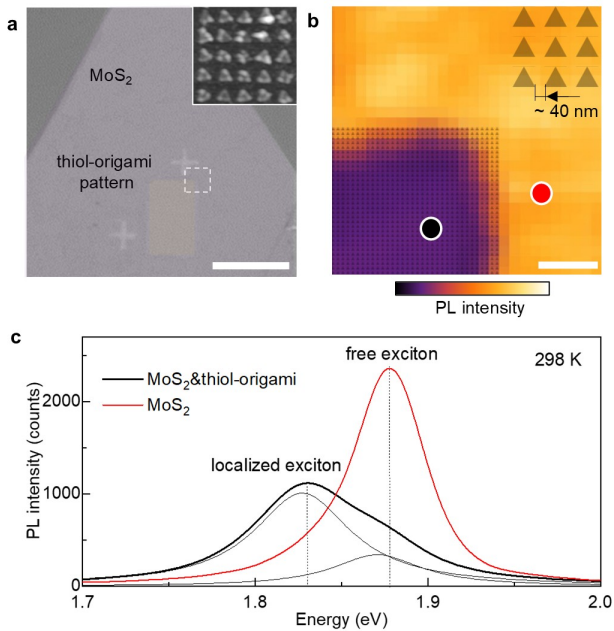

**Figure 2. Optical properties of MoS$_2$ flake on thiol–origami pattern.** (**a**) Bright-field microscopy image of the MoS$_2$ flake (shaded in purple) transferred onto the thiol–origami pattern (shaded in yellow). The nanopatterned area is 20 μm x 25 μm and consists of arrays of triangular binding sites with a size of 120 nm, arranged in a square lattice with a 170 nm period. The inset shows typical dry-mode AFM images of Si/SiO$_2$ chips with DNA triangles bearing 18 19 nt thymine "linker" extensions, each terminated with a thiol group at its 5' end, arranged with a 170 nm period between triangles. (**b**) Room-temperature PL maps, measured in the area of the sample highlighted in panel (**a**). PL excitation energy is 2.21 eV (560 nm). Each gray triangle indicates the position of a DNA triangle. The inset shows a schematic of the patterned area. (**c**) Room-temperature PL spectra of two different spots of the PL map presented in panel **b**. Black spot: thiol–origami-MoS$_2$ hybrid assembly, deconvoluted PL spectrum. Red spot: MoS$_2$. Scale bars: 30 μm (panel **a**), and 5 μm (panel **b**).

## Single photon emission

To optically resolve PL signal variations directly corresponding to the positions of individual thiol–origami structures, we patterned MoS$_2$ with a square array of DNA triangles with a 1000 nm period (Fig. 3a) and recorded a PL map. Fig. 3b depicts the ratio of PL intensities for the localized and free exciton peaks at 1.83 eV and 1.88 eV, respectively. The emission intensity of the free exciton is significantly reduced at all lattice sites, indicating efficient exciton capture or funneling into localized emitters. This confirms that the thiolated strands on each individual DNA triangle bind to MoS$_2$, forming trapping sites for excitons and resulting in emission at an energy 50 meV lower than that of the free exciton in MoS$_2$.

Next, we investigated the emission properties of thiol-functionalized MoS$_2$ at 4 K. Fig. 3c depicts a typical PL spectrum from an individual site, marked by the red circle in Fig. 3b. At low temperature, both free and localized MoS$_2$ emission peaks are blue-shifted by ∼ 50 meV.[72, 73] The broad PL of localized excitons resolves into several sharp peaks, each with a linewidth below 1 meV. Fig. 3d presents the second-order photon-correlation function $g^{(2)}(\tau)$ for a single sharp peak (marked in grey in Fig. 3c). The measured second-order correlation function exhibits a prominent dip at zero-time delay with $g^{(2)}(\tau) = 0.31$. The value of $g^{(2)}(0)$, well below the 0.5 threshold, demonstrates single-photon emission from our thiol–origami patterned $MoS_2$. Time-resolved PL measurements reveal that the PL lifetimes of the SPE are typically on the order of nanoseconds (Fig. 3e). This experimental observation is in stark contrast to helium-ion-induced emitters in $MoS_2$, which exhibit microsecond lifetimes,[11] and aligns more closely with the SPEs in $WSe_2$, which have nanosecond lifetimes.[67, 68, 69] Remarkably, the quantum emitters demonstrate minimal photobleaching, blinking, and spectral diffusion, indicating robust performance (Fig. 3f and Supplementary Fig. 5).

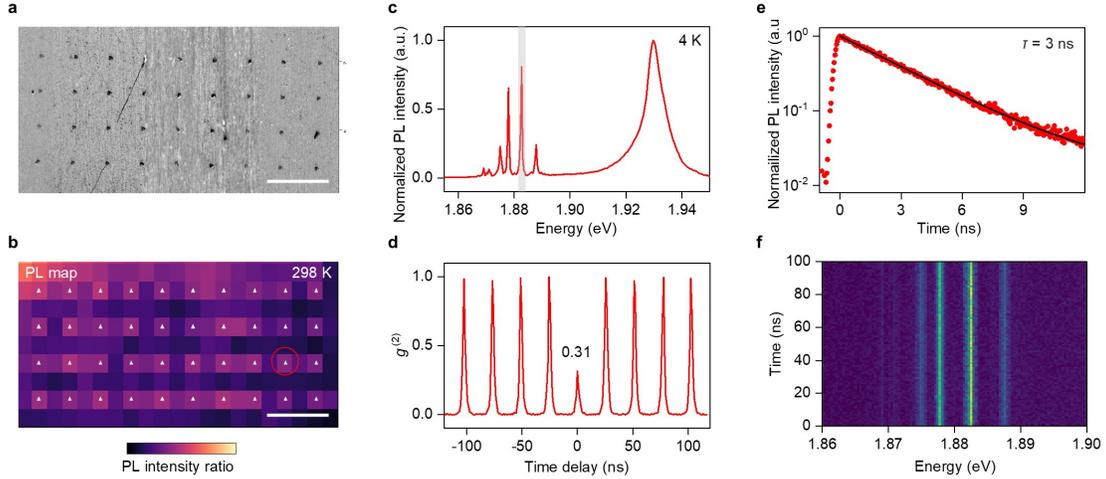

**Figure 3. DNA origami-programmable single-photon emission in $MoS_2$–thiol molecule hybrids. (a)** Typical dry-mode AFM images of $Si/SiO_2$ chips with DNA triangles bearing 18 19-nt thymine "linker" extensions, each terminated with a thiol group at its 5' end, arranged with a 1000 nm period between triangles. **(b)** Map of the ratio of localized to free exciton PL peaks in a thiol–origami-$MoS_2$ hybrid assembly. Each white triangle indicates the position of a DNA triangle. **(c)** Low-temperature PL spectrum from the spot indicated by the red circle in panel (**b**). **(d)** Correlation function of the localized defect state marked by the gray bar in panel (**c**), showing second-order coherence at zero time delay with $g^{(2)}(0) = 0.31$. **(e)** Time-resolved PL of localized defect state marked by the gray bar in panel (**c**) showing a single-exponential decay with a decay time of $3.099 \pm 0.001$ ns. **(f)** Low-temperature PL intensity of localized states as a function of measurement time. Scale bars: 2 µm (panels **a** and **b**).

Furthermore, we measured low-temperature PL spectra for 33 thiol–origami binding sites (shown in Supplementary Fig. 4). We observed localized emission at all 33 positions, corresponding to the thiol–origami locations. Representative PL spectra from five of these positions are shown in Supplementary Fig. 4. Each spectrum exhibits four to six emission lines that are absent in non-patterned areas. This demonstrates that patterning with thiol–origami enables the deterministic induction of quantum defects in atomically thin $MoS_2$. All measured defect emitters exhibited low-temperature localized emission red-shifted by ~50 meV from the free exciton peak, consistent with the room-temperature emission. Out of the 33 measured thiol–origami sites, 29 exhibited single-photon emission peaks, resulting in a single-photon emitter placement yield of 0.88 (see Supplementary Fig. 4 for PL data and statistical analysis). The positioning accuracy of our technique is predominantly determined by the positioning accuracy of DNA origami placement, which is ~10 nm, as previously demonstrated.[52, 54] Near-field optical techniques can be further used to reveal the spatial accuracy of thiol-induced quantum emitters.

### Functionalization mechanism and scalability

It is commonly assumed that thiol-derivative molecules bind to sulfur vacancies in $MoS_2$.[60, 74, 75, 76] Consistent with this assumption, we attribute the binding to chemisorption of thiols to sulfur vacancies in $MoS_2$, as illustrated in Fig. 4a. Reported sulfur vacancy densities in $MoS_2$ vary significantly depending on the experimental conditions and evaluation technique, ranging from approximately $10^{13}$ cm$^{-2}$ as measured by super-resolved optical mapping[60] to $10^{15}$ cm$^{-2}$ as reported by TEM and STM investigations.[77, 78, 79] However, the relatively high density of these vacancies compared to our thiolated strands (estimated at ~$10^{11}$ cm$^{-2}$, with each DNA triangle carrying 18 thiolated strands) still ensures efficient binding.

To assess the scalability of our approach, we fabricated a matrix of 4750 individual thiol–origami arranged in a square array with periods decreasing gradually from 1000 nm to 170 nm (Fig. 4b; see Supplementary Fig. 3 for AFM characterization). Due to the diffraction-limited optical resolution of ~1 μm, we could not spatially resolve PL signals from individual lattice sites when the thiol–origami periods ranged from 170 to 800 nm; instead, we observed relatively homogeneous PL signals in these regions. PL maps of free and localized excitons are shown in Figs. 4c and 4d, respectively. The highest-density pattern with 170 nm period exhibits near-complete quenching of free exciton emission and strong localized exciton emission (Figs. 4e).

We calculated the average intensity of localized and free exciton emission for each step of the pattern. Figure 4f shows the ratio of localized to free exciton emission as a function of DNA triangle density. Our experimental values are well fitted by a linear relationship (see Supplementary Note 3). Based on this observed linear dependence, we propose that the single-photon emitter placement yield could be independent of the pattern period and may reach 88%, even for the densest patterns. Then, using our densest patterns, up to $5 \times 10^5$ single-photon light sources could be easily achieved on a single 1000 μm$^2$ $MoS_2$ equilateral triangle with a side length of 50 μm, demonstrating the remarkable scalability of our technique.

Overall, our DNA origami-programmed functionalization of $MoS_2$ is site-selective, non-destructive, and uses inexpensive materials that are easy and safe to handle. By employing our functionalization approach, we were able to introduce quantum emitters in $MoS_2$ with molecular-level precision, previously achievable only with destructive top-down techniques such as focused-ion beam irradiation. These results highlight DNA origami as a unique and versatile tool for customizing $MoS_2$ materials with precisely targeted localized quantum defects acting as single-photon emitters. Further tuning of the number of thiol molecules per origami and exploring different organic molecules provides a route to improve single-photon purity[80] and generate chiral quantum light.[81]

### CONCLUSIONS

In summary, we tuned the optical properties of monolayer $MoS_2$ via functionalization with thiol molecules, precisely positioned on chip surfaces using a DNA origami placement technique. After transferring the $MoS_2$ monolayer onto micron-scale thiol–origami patterns, we observed single-photon emission, attributed to exciton trapping caused by thiol-$MoS_2$ binding. Unprecedented control over the density of these localized excitons was achieved by adjusting the square-lattice period of the thiol–origami pattern. Looking forward, our technique can be used for precise functionalization of a wide range of 2D materials, including TMDs, graphene, and other 2D van der Waals structures. In nanophotonics, our hybrid 2D organic-inorganic structures will enable exciton landscape engineering, where DNA origami can precisely control the number and density of localized defects for next-generation circuits and quantum emitters[11] for quantum communications. In addition to nanophotonic devices, our hybrid approach may be useful for nanoelectronic applications[82] and any heterogeneous fabrication process requiring the integration of molecules or nanoparticles with 2D materials at high spatial accuracy and orientation control.

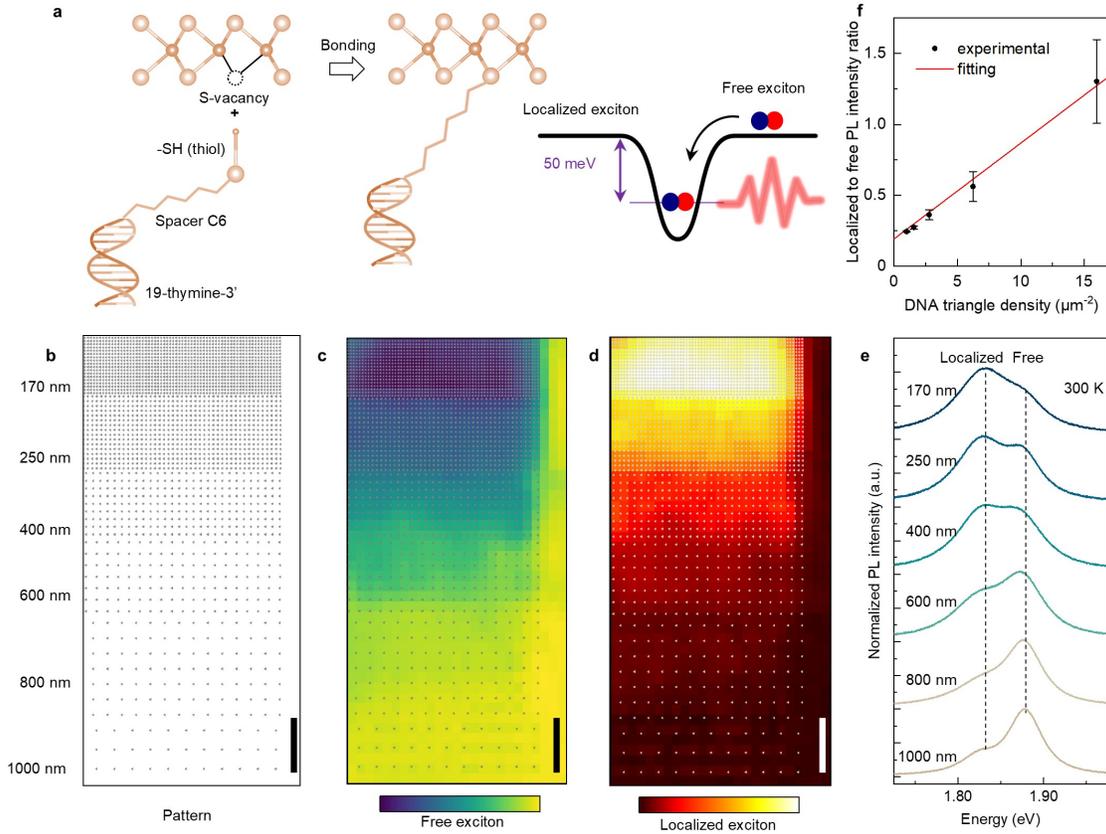

**Figure 4. DNA origami-programmable exciton landscape in thiol-MoS₂ hybrids.** (**a**) Schematic illustration of the exciton localization scheme in thiol-MoS₂ hybrids. (**b**) Schematic of the patterned area. The nanopatterned area is 20 μm x 25 μm and consists of arrays of triangular binding sites with a size of 120 nm in a square lattice with periods ranging from 1000 nm to 170 nm. (**c, d**) Free exciton and localized exciton PL intensity maps, measured at 1.88 eV and 1.83 eV, respectively. Each gray triangle indicates the position of a DNA triangle. (**e**) Representative room-temperature PL spectra of each step in the gradient pattern presented in panel (**b**). (**f**) The ratio of PL intensities of localized to free excitons versus DNA triangle density. Dots represent experimental values of ratios of averaged localized and free PL intensity, calculated for each step area in panels (**c**) and (**d**). The red line represents a linear fit (see Supplementary Note 3). Scale bars: 4 μm (panels **b**, **c**, and **d**).

## ACKNOWLEDGMENTS

We thank Philipp Altpeter and Christian Obermayer for assistance in the clean room. I.M. and T.L. acknowledge funding from the ERC consolidator grant "DNA Funs" (Project ID: 818635) T.L. further acknowledges support from the cluster of excellence e-conversion EXC 2089/1- 390776260. This work was funded by the Federal Ministry of Education and Research (BMBF) and the Free State of Bavaria under the Excellence Strategy of the Federal Government and the Länder through the ONE MUNICH Project Munich Multiscale Biofabrication. K.W. and T.T. acknowledge support from the JSPS KAKENHI (Grant Numbers 21H05233 and 23H02052), the CREST (JPMJCR24A5), JST and World Premier International Research Center Initiative (WPI), MEXT, Japan. A.H. acknowledges funding by the Deutsche Forschungsgemeinschaft (DFG, German Research Foundation) within the

AUTHOR CONTRIBUTIONS STATEMENT

I.M., A.B. and A.H. designed this study. I.M. designed, assembled, and purified DNA origami samples, performed DNA origami placement experiments, and conducted AFM measurements and data analysis with assistance from E.E., C.S., and A.A. Z.L. performed MoS$_2$ monolayer synthesis and, together with I.M., transferred it onto the origami patterns. K.W. and T.T. performed h-BN synthesis. S.Z. performed optical spectroscopy. I.M., A.B., and S.Z. analyzed the data and wrote the manuscript with input from A.H. and T.L.

COMPETING INTERESTS STATEMENT

The authors declare no competing financial interest.

REFERENCES

1.  Sutula M, Christen I, Bersin E, Walsh MP, Chen KC, Mallek J, *et al.* Large-scale optical characterization of solid-state quantum emitters. *Nature Materials* 2023, **22**(11): 1338-1344.

2.  Aharonovich I, Englund D, Toth M. Solid-state single-photon emitters. *Nature Photonics* 2016, **10**(10): 631-641.

3.  Novoselov KS, Jiang D, Schedin F, Booth TJ, Khotkevich VV, Morozov SV, *et al.* Two-dimensional atomic crystals. *Proceedings of the National Academy of Sciences* 2005, **102**(30): 10451-10453.

4.  Wang QH, Kalantar-Zadeh K, Kis A, Coleman JN, Strano MS. Electronics and optoelectronics of two-dimensional transition metal dichalcogenides. *Nature Nanotechnology* 2012, **7**(11): 699-712.

5.  Michaelis de Vasconcellos S, Wigger D, Wurstbauer U, Holleitner AW, Bratschitsch R, Kuhn T. Single-Photon Emitters in Layered Van der Waals Materials. *physica status solidi (b)* 2022, **259**(4): 2100566.

6.  Sortino L, Zotev PG, Phillips CL, Brash AJ, Cambiasso J, Marensi E, *et al.* Bright single photon emitters with enhanced quantum efficiency in a two-dimensional semiconductor coupled with dielectric nano-antennas. *Nature Communications* 2021, **12**(1): 6063.

7.  Palacios-Berraquero C, Kara DM, Montblanch ARP, Barbone M, Latawiec P, Yoon D, *et al.* Large-scale quantum-emitter arrays in atomically thin semiconductors. *Nature Communications* 2017, **8**(1): 15093.

8.  Branny A, Kumar S, Proux R, Gerardot BD. Deterministic strain-induced arrays of quantum emitters in a two-dimensional semiconductor. *Nature Communications* 2017, **8**(1): 15053.

9.  Abramov AN, Chestnov IY, Alimova ES, Ivanova T, Mukhin IS, Krizhanovskii DN, *et al.* Photoluminescence imaging of single photon emitters within nanoscale strain profiles in monolayer WSe2. *Nature Communications* 2023, **14**(1): 5737.

10. Parto K, Azzam SI, Banerjee K, Moody G. Defect and strain engineering of monolayer WSe2 enables site-controlled single-photon emission up to 150 K. *Nature Communications* 2021, **12**(1): 3585.

11. Klein J, Lorke M, Florian M, Sigger F, Sigl L, Rey S*, et al.* Site-selectively generated photon emitters in monolayer MoS2 via local helium ion irradiation. *Nature Communications* 2019, **10**(1): 2755.

12. Yanev ES, Darlington TP, Ladyzhets SA, Strasbourg MC, Trovatello C, Liu S*, et al.* Programmable nanowrinkle-induced room-temperature exciton localization in monolayer WSe2. *Nature Communications* 2024, **15**(1): 1543.

13. So J-P, Kim H-R, Baek H, Jeong K-Y, Lee H-C, Huh W*, et al.* Electrically driven strain-induced deterministic single-photon emitters in a van der Waals heterostructure. *Science Advances* 2021, **7**(43): eabj3176.

14. Barthelmi K, Klein J, Hötger A, Sigl L, Sigger F, Mitterreiter E*, et al.* Atomistic defects as single-photon emitters in atomically thin MoS2. *Appl Phys Lett* 2020, **117**(7).

15. Piao Y, Meany B, Powell LR, Valley N, Kwon H, Schatz GC*, et al.* Brightening of carbon nanotube photoluminescence through the incorporation of sp3 defects. *Nature Chemistry* 2013, **5**(10): 840-845.

16. Ma X, Hartmann NF, Baldwin JKS, Doorn SK, Htoon H. Room-temperature single-photon generation from solitary dopants of carbon nanotubes. *Nature Nanotechnology* 2015, **10**(8): 671-675.

17. Huang Y, Zheng W, Qiu Y, Hu P. Effects of Organic Molecules with Different Structures and Absorption Bandwidth on Modulating Photoresponse of MoS2 Photodetector. *ACS Applied Materials & Interfaces* 2016, **8**(35): 23362-23370.

18. Huang Y, Zhuge F, Hou J, Lv L, Luo P, Zhou N*, et al.* Van der Waals Coupled Organic Molecules with Monolayer MoS2 for Fast Response Photodetectors with Gate-Tunable Responsivity. *ACS nano* 2018, **12**(4): 4062-4073.

19. Jariwala D, Howell SL, Chen K-S, Kang J, Sangwan VK, Filippone SA*, et al.* Hybrid, Gate-Tunable, van der Waals p–n Heterojunctions from Pentacene and MoS2. *Nano Lett* 2016, **16**(1): 497-503.

20. Chen R, Lin C, Yu H, Tang Y, Song C, Yuwen L*, et al.* Templating C60 on MoS2 Nanosheets for 2D Hybrid van der Waals p–n Nanoheterojunctions. *Chem Mater* 2016, **28**(12): 4300-4306.

21. Bettis Homan S, Sangwan VK, Balla I, Bergeron H, Weiss EA, Hersam MC. Ultrafast Exciton Dissociation and Long-Lived Charge Separation in a Photovoltaic Pentacene–MoS2 van der Waals Heterojunction. *Nano Lett* 2017, **17**(1): 164-169.

22. Kim J-S, Yoo H-W, Choi HO, Jung H-T. Tunable Volatile Organic Compounds Sensor by Using Thiolated Ligand Conjugation on MoS2. *Nano Lett* 2014, **14**(10): 5941-5947.

23.     Szalai AM, Ferrari G, Richter L, Hartmann J, Kesici M-Z, Ji B, *et al.* Single-molecule dynamic structural biology with vertically arranged DNA on a fluorescence microscope. *Nature Methods* 2025, **22**(1): 135-144.

24.     Li BL, Luo HQ, Lei JL, Li NB. Hemin-functionalized MoS2 nanosheets: enhanced peroxidase-like catalytic activity with a steady state in aqueous solution. *RSC Advances* 2014, **4**(46): 24256-24262.

25.     Zhu T, Yuan L, Zhao Y, Zhou M, Wan Y, Mei J, *et al.* Highly mobile charge-transfer excitons in two-dimensional $WS_2$/tetracene heterostructures. *Science Advances* 2018, **4**(1): eaao3104.

26.     Thompson JJP, Gerhard M, Witte G, Malic E. Optical signatures of Förster-induced energy transfer in organic/TMD heterostructures. *npj 2D Materials and Applications* 2023, **7**(1): 69.

27.     Amani M, Lien D-H, Kiriya D, Xiao J, Azcatl A, Noh J, *et al.* Near-unity photoluminescence quantum yield in $MoS_2$. *Science* 2015, **350**(6264): 1065-1068.

28.     Wang S, Yang X, Hou L, Cui X, Zheng X, Zheng J. Organic covalent modification to improve thermoelectric properties of TaS2. *Nature Communications* 2022, **13**(1): 4401.

29.     Varade V, Haider G, Slobodeniuk A, Korytar R, Novotny T, Holy V, *et al.* Chiral Light Emission from a Hybrid Magnetic Molecule–Monolayer Transition Metal Dichalcogenide Heterostructure. *ACS nano* 2023, **17**(3): 2170-2181.

30.     Schmidt H, Giustiniano F, Eda G. Electronic transport properties of transition metal dichalcogenide field-effect devices: surface and interface effects. *Chem Soc Rev* 2015, **44**(21): 7715-7736.

31.     Chen X, Bartlam C, Lloret V, Moses Badlyan N, Wolff S, Gillen R, *et al.* Covalent Bisfunctionalization of Two-Dimensional Molybdenum Disulfide. *Angew Chem Int Ed* 2021, **60**(24): 13484-13492.

32.     Wei T, Al-Fogra S, Hauke F, Hirsch A. Direct Laser Writing on Graphene with Unprecedented Efficiency of Covalent Two-Dimensional Functionalization. *J Am Chem Soc* 2020, **142**(52): 21926-21931.

33.     Kollipara PS, Li J, Zheng Y. Optical Patterning of Two-Dimensional Materials. *Research* 2020, **2020**.

34.     Tan X, Wang S, Zhang Q, He J, Chen S, Qu Y, *et al.* Laser doping of 2D material for precise energy band design. *Nanoscale* 2023, **15**(21): 9297-9303.

35.     Rothemund PWK. Folding DNA to create nanoscale shapes and patterns. *Nature* 2006, **440**(7082): 297-302.

36.     Douglas SM, Dietz H, Liedl T, Högberg B, Graf F, Shih WM. Self-assembly of DNA into nanoscale three-dimensional shapes. *Nature* 2009, **459**(7245): 414-418.

37.     Seeman NC. DNA in a material world. *Nature* 2003, **421**(6921): 427-431.

38.     Yan H, Park SH, Finkelstein G, Reif JH, LaBean TH. DNA-templated self-assembly of protein arrays and highly conductive nanowires. *Science* 2003, **301**(5641)**:** 1882-1884.

39.     Aldaye FA, Palmer AL, Sleiman HF. Assembling materials with DNA as the guide. *Science* 2008, **321**(5897)**:** 1795-1799.

40.     Wang P, Huh J-H, Lee J, Kim K, Park KJ, Lee S*, et al.* Magnetic Plasmon Networks Programmed by Molecular Self-Assembly. *Adv Mater* 2019, **31**(29)**:** 1901364.

41.     Liu X, Zhang F, Jing X, Pan M, Liu P, Li W*, et al.* Complex silica composite nanomaterials templated with DNA origami. *Nature* 2018, **559**(7715)**:** 593-598.

42.     Sun W, Boulais E, Hakobyan Y, Wang WL, Guan A, Bathe M*, et al.* Casting inorganic structures with DNA molds. *Science* 2014, **346**(6210)**:** 1258361.

43.     Kolbeck PJ, Dass M, Martynenko IV, van Dijk-Moes RJ, Brouwer KJ, van Blaaderen A*, et al.* A DNA origami fiducial for accurate 3D AFM imaging. *bioRxiv* 2022.

44.     Maune HT, Han S-p, Barish RD, Bockrath M, Iii WAG, Rothemund PWK*, et al.* Self-assembly of carbon nanotubes into two-dimensional geometries using DNA origami templates. *Nature Nanotechnology* 2010, **5**(1)**:** 61-66.

45.     Kuzyk A, Schreiber R, Fan Z, Pardatscher G, Roller E-M, Högele A*, et al.* DNA-based self-assembly of chiral plasmonic nanostructures with tailored optical response. *Nature* 2012, **483**(7389)**:** 311-314.

46.     Voigt NV, Tørring T, Rotaru A, Jacobsen MF, Ravnsbæk JB, Subramani R*, et al.* Single-molecule chemical reactions on DNA origami. *Nature Nanotechnology* 2010, **5**(3)**:** 200-203.

47.     Knudsen JB, Liu L, Bank Kodal AL, Madsen M, Li Q, Song J*, et al.* Routing of individual polymers in designed patterns. *Nature Nanotechnology* 2015, **10**(10)**:** 892-898.

48.     Hartl C, Frank K, Amenitsch H, Fischer S, Liedl T, Nickel B. Position Accuracy of Gold Nanoparticles on DNA Origami Structures Studied with Small-Angle X-ray Scattering. *Nano Lett* 2018, **18**(4)**:** 2609-2615.

49.     Funke JJ, Dietz H. Placing molecules with Bohr radius resolution using DNA origami. *Nature Nanotechnology* 2016, **11**(1)**:** 47-52.

50.     Kershner RJ, Bozano LD, Micheel CM, Hung AM, Fornof AR, Cha JN*, et al.* Placement and orientation of individual DNA shapes on lithographically patterned surfaces. *Nature Nanotechnology* 2009, **4**(9)**:** 557-561.

51.     Hung AM, Micheel CM, Bozano LD, Osterbur LW, Wallraff GM, Cha JN. Large-area spatially ordered arrays of gold nanoparticles directed by lithographically confined DNA origami. *Nature Nanotechnology* 2010, **5**(2)**:** 121-126.

52.     Sikeler C, Haslinger F, Martynenko IV, Liedl T. DNA Origami-Directed Self-Assembly of Gold Nanospheres for Plasmonic Metasurfaces. *Adv Funct Mater* 2024, **34**(42)**:** 2404766.

53. Martynenko IV, Erber E, Ruider V, Dass M, Posnjak G, Yin X, *et al.* Site-directed placement of three-dimensional DNA origami. *Nature Nanotechnology* 2023, **18**(12): 1456-1462.

54. Gopinath A, Miyazono E, Faraon A, Rothemund PWK. Engineering and mapping nanocavity emission via precision placement of DNA origami. *Nature* 2016, **535**(7612)**:** 401-405.

55. Gopinath A, Thachuk C, Mitskovets A, Atwater HA, Kirkpatrick D, Rothemund PWK. Absolute and arbitrary orientation of single-molecule shapes. *Science* 2021, **371**(6531)**:** eabd6179.

56. Korlach J, Marks PJ, Cicero RL, Gray JJ, Murphy DL, Roitman DB*, et al.* Selective aluminum passivation for targeted immobilization of single DNA polymerase molecules in zero-mode waveguide nanostructures. *Proceedings of the National Academy of Sciences* 2008, **105**(4)**:** 1176-1181.

57. Beer NR, Hindson BJ, Wheeler EK, Hall SB, Rose KA, Kennedy IM, *et al.* On-Chip, Real-Time, Single-Copy Polymerase Chain Reaction in Picoliter Droplets. *Anal Chem* 2007, **79**(22)**:** 8471-8475.

58. Simon JR, Maksimov D, Lotze C, Wiechers P, Felipe JPG, Kobin B*, et al.* Atomic-scale perspective on individual thiol-terminated molecules anchored to single S vacancies in $\mathrm{MoS}_{2}$. *Phys Rev B* 2024, **110**(4)**:** 045407.

59. Makarova M, Okawa Y, Aono M. Selective Adsorption of Thiol Molecules at Sulfur Vacancies on MoS2(0001), Followed by Vacancy Repair via S–C Dissociation. *J Phys Chem C* 2012, **116**(42)**:** 22411-22416.

60. Zhang M, Lihter M, Chen T-H, Macha M, Rayabharam A, Banjac K*, et al.* Super-resolved Optical Mapping of Reactive Sulfur-Vacancies in Two-Dimensional Transition Metal Dichalcogenides. *ACS nano* 2021, **15**(4)**:** 7168-7178.

61. Gopinath A, Rothemund PWK. Optimized Assembly and Covalent Coupling of Single-Molecule DNA Origami Nanoarrays. *ACS nano* 2014, **8**(12)**:** 12030-12040.

62. Douglas SM, Marblestone AH, Teerapittayanon S, Vazquez A, Church GM, Shih WM. Rapid prototyping of 3D DNA-origami shapes with caDNAno. *Nucleic Acids Res* 2009, **37**(15)**:** 5001-5006.

63. Kim D-N, Kilchherr F, Dietz H, Bathe M. Quantitative prediction of 3D solution shape and flexibility of nucleic acid nanostructures. *Nucleic Acids Res* 2011, **40**(7)**:** 2862-2868.

64. Mak KF, Lee C, Hone J, Shan J, Heinz TF. Atomically Thin $\mathrm{MoS}_{2}$: A New Direct-Gap Semiconductor. *Phys Rev Lett* 2010, **105**(13)**:** 136805.

65. Neumann A, Lindlau J, Nutz M, Mohite AD, Yamaguchi H, Högele A. Signatures of defect-localized charged excitons in the photoluminescence of monolayer molybdenum disulfide. *Physical Review Materials* 2018, **2**(12)**:** 124003.

66. Mitterreiter E, Schuler B, Micevic A, Hernangómez-Pérez D, Barthelmi K, Cochrane KA*, et al.* The role of chalcogen vacancies for atomic defect emission in MoS2. *Nature Communications* 2021, **12**(1)**:** 3822.

67. Tonndorf P, Schmidt R, Schneider R, Kern J, Buscema M, Steele GA, *et al.* Single-photon emission from localized excitons in an atomically thin semiconductor. *Optica* 2015, **2**(4): 347-352.

68. Srivastava A, Sidler M, Allain AV, Lembke DS, Kis A, Imamoğlu A. Optically active quantum dots in monolayer WSe2. *Nature Nanotechnology* 2015, **10**(6): 491-496.

69. He Y-M, Clark G, Schaibley JR, He Y, Chen M-C, Wei Y-J, *et al.* Single quantum emitters in monolayer semiconductors. *Nature Nanotechnology* 2015, **10**(6): 497-502.

70. Koperski M, Nogajewski K, Arora A, Cherkez V, Mallet P, Veuillen JY, *et al.* Single photon emitters in exfoliated WSe2 structures. *Nature Nanotechnology* 2015, **10**(6): 503-506.

71. Chakraborty C, Kinnischtzke L, Goodfellow KM, Beams R, Vamivakas AN. Voltage-controlled quantum light from an atomically thin semiconductor. *Nature Nanotechnology* 2015, **10**(6): 507-511.

72. Christopher JW, Goldberg BB, Swan AK. Long tailed trions in monolayer MoS2: Temperature dependent asymmetry and resulting red-shift of trion photoluminescence spectra. *Sci Rep* 2017, **7**(1): 14062.

73. Tongay S, Zhou J, Ataca C, Lo K, Matthews TS, Li J, *et al.* Thermally Driven Crossover from Indirect toward Direct Bandgap in 2D Semiconductors: MoSe2 versus MoS2. *Nano Lett* 2012, **12**(11): 5576-5580.

74. Kirubasankar B, Won YS, Adofo LA, Choi SH, Kim SM, Kim KK. Atomic and structural modifications of two-dimensional transition metal dichalcogenides for various advanced applications. *Chem Sci* 2022, **13**(26): 7707-7738.

75. Mouri S, Miyauchi Y, Matsuda K. Tunable Photoluminescence of Monolayer MoS2 via Chemical Doping. *Nano Lett* 2013, **13**(12): 5944-5948.

76. Sim DM, Kim M, Yim S, Choi M-J, Choi J, Yoo S, *et al.* Controlled Doping of Vacancy-Containing Few-Layer MoS2 via Highly Stable Thiol-Based Molecular Chemisorption. *ACS nano* 2015, **9**(12): 12115-12123.

77. Li Z, Bretscher H, Rao A. Chemical passivation of 2D transition metal dichalcogenides: strategies, mechanisms, and prospects for optoelectronic applications. *Nanoscale* 2024, **16**(20): 9728-9741.

78. Lin J, Pantelides ST, Zhou W. Vacancy-Induced Formation and Growth of Inversion Domains in Transition-Metal Dichalcogenide Monolayer. *ACS nano* 2015, **9**(5): 5189-5197.

79. Liu S, Liu Y, Holtzman L, Li B, Holbrook M, Pack J, *et al.* Two-Step Flux Synthesis of Ultrapure Transition-Metal Dichalcogenides. *ACS nano* 2023, **17**(17): 16587-16596.

80. He X, Htoon H, Doorn SK, Pernice WHP, Pyatkov F, Krupke R, *et al.* Carbon nanotubes as emerging quantum-light sources. *Nature Materials* 2018, **17**(8): 663-670.

81. Li X, Jones AC, Choi J, Zhao H, Chandrasekaran V, Pettes MT, *et al.* Proximity-induced chiral quantum light generation in strain-engineered WSe2/NiPS3 heterostructures. *Nature Materials* 2023, **22**(11): 1311-1316.

82. Markeev PA, Najafidehaghani E, Samu GF, Sarosi K, Kalkan SB, Gan Z, *et al.* Exciton Dynamics in MoS2-Pentacene and WSe2-Pentacene Heterojunctions. *ACS nano* 2022, **16**(10): 16668-16676.

83. Bilgin I, Raeliarijaona AS, Lucking MC, Hodge SC, Mohite AD, de Luna Bugallo A, *et al.* Resonant Raman and Exciton Coupling in High-Quality Single Crystals of Atomically Thin Molybdenum Diselenide Grown by Vapor-Phase Chalcogenization. *ACS nano* 2018, **12**(1): 740-750.

84. Pizzocchero F, Gammelgaard L, Jessen BS, Caridad JM, Wang L, Hone J, *et al.* The hot pick-up technique for batch assembly of van der Waals heterostructures. *Nature Communications* 2016, **7**(1): 11894.

## METHODS

### DNA Origami design, preparation and purification

The 'sameside sharp triangle' design is adapted from Ref.[1] We modified 18 staples from the original design by extending them with 20-nucleotide adenine extensions on the 3′ end. The positions of these extended staples are shown in Supplementary Fig. S1. These 20-nucleotide adenine-extended staples serve as linkers that bind to a 19-nucleotide thymine strand bearing a single thiol molecule on its 5′ end in a 'shear' configuration.

DNA origami triangles were folded by mixing scaffold strands (7249 nucleotides long M13mp18 single-stranded DNA) with an excess of staple strands in folding buffer (10 mM Tris, 1 mM EDTA, 12.5 mM $MgCl_2$, pH 8.35). The scaffold DNA was mixed with all staples except for 18 staples, which were replaced with modified versions containing 20-nucleotide adenine extensions at the 3′ ends (IDT Technologies, 200 μM). The final concentrations in the folding buffer were: 20 nM scaffold strand, 100 nM staple strands, and 2000 nM polyA-modified staples (IDT Technologies, 200 μM). The samples were annealed in a PCR machine (Biometra TRIO Thermal Cycler, Analytik Jena) and purified from excess staples by Amicon filtration as described in Ref.[53] After Amicon purification, the concentration of triangles typically ranged from 70 to 100 nM, with a recovery yield of 40-50%. Scaffold strands were produced from M13 phage replication in Escherichia coli. All chemicals were obtained from Sigma Aldrich unless otherwise stated.

### Preparation of the substrates and DNA origami placement

Si/SiO$_2$ substrates were patterned using electron-beam lithography, following protocols from Ref.[2] with slight modifications. A 4-inch Si/SiO$_2$ wafer with a 100 nm thermal oxide layer (Microchemicals) was diced into 1 cm × 1 cm chips. Clean chips were primed with 10 mL of hexamethyldisilazane (HMDS) in a 4 L desiccator. The priming time was optimized to maintain a Si/SiO$_2$ surface contact angle of 70°-75° after HMDS deposition. Triangular binding sites, where individual DNA triangles would bind, were patterned into poly(methyl methacrylate) resist by electron-beam lithography. Similarly, we patterned approximately 1-μm lines to which multiple origami would bind at random positions and orientations. The chips were then developed with a 1:3 solution of methyl isobutyl ketone (MIBK) and isopropanol (IPA). The HMDS in the developed areas was removed using O$_2$ plasma for 6 seconds in a plasma cleaner (PICO). The resist was stripped by ultrasonication in N-methyl pyrrolidone (NMP) at 50°C for 30 minutes. The substrates were briefly rinsed with 2-isopropanol, dried in a nitrogen stream, and used immediately.

DNA triangles were bound to the patterned substrates as previously described in Refs. [53, 61] A 20-60 μL drop of freshly folded and purified DNA origami was deposited onto the surface of the chips in placement buffer (5 mM Tris, 35 mM MgCl₂, pH 8.35). After incubation for 1 hour at RT in a 100% humidity incubator, excess DNA origami was removed from the surface by performing eight buffer replacement steps and purifying with 0.1% Tween 20.

### Post-placement functionalization of DNA origami

After placement, DNA triangles were labelled with a thiol group and dried using ethanol drying. Functionalization was performed by incubating the placed origami, bearing 21-nucleotide adenine extensions, in a placement buffer containing 200 nM of a 19-nucleotide thymine strand bearing a single thiol group (Biomers, HPLC purified, 100 μM) for 1 hour at room temperature. Excess ssDNA strands were removed from the surface by performing eight buffer replacement steps and purifying with 0.1% Tween 20. After this step, chips with absorbed DNA origami were air-dried using an ethanol dilution series as previously described.[61]

### MoS₂ monolayer synthesis and transfer

Monolayers of MoS₂ were synthesized by the vapor phase chalcogenization method[83] on thermally oxidized SiO₂/Si substrates with oxide thickness of 285 nm. A three-zone furnace system (Carbolite Gero) equipped with a 1 inch quartz tube was used for growth under ambient pressure. In order to precisely place MoS₂ monolayers onto molecule-origami patterns, a dry- transfer method with PDMS/PC stamps was used.[84] An hBN flake with a thickness around 100 nm was firstly picked up at 50 °C, followed by a triangular monolayer MoS₂ at 145 °C. The entire stack was subsequently released from the stamp onto a SiO₂/Si target substrate with molecule-origami pattern at a temperature of 180°C. The sample was then soaked in chloroform solution for 5 min to remove polycarbonate residues, cleaned by acetone and isopropanol and annealed under ultrahigh vacuum for 12 hours.

### Characterization techniques and data analysis

UV-vis absorption measurements were performed with a NanoDrop ND-1000 Spectrophotometer (Thermo Scientific). Tapping-mode AFM of dried Si/SiO₂ substrates with triangular DNA origami was carried out on a Dimension ICON AFM (Bruker). OTESPA silicon tips (300 kHz, Vecco Probes) were used for imaging in air. Images are analysed with the Software Gwyddeon.

Hyperspectral Photoluminescence (PL) imaging was performed using a custom-built scanning confocal microscope. The sample was mounted on piezo-stepping units (ANPxyz101, attocube system) for precise positioning with respect to the confocal spot of an apochromatic objective (LT-APO/532-RAMAN, attocube system). Continuous-wave diode lasers with a power of 2 μW were used for excitation to prevent sample damage. Spectrally sharp short-pass and long-pass filters (Semrock) were employed in the excitation and detection paths, respectively, to eliminate laser light. The sample's PL signal was dispersed by a monochromator (Acton SpectraPro 300i, Roper Scientific) with a 300 grooves/mm grating and detected using a Peltier-cooled charge-coupled device (CCD, Andor iDus 416). We subtracted the hBN background emission from the raw PL data.

The low-temperature PL measurements were performed in a close-cycle cryostat (attoDRY800, attocube system) with a base temperature of 4 K, using the same scanning confocal microscope as for the room-temperature measurements. The second-order photon-correlation $g^{(2)}(\tau)$ measurements were carried out using a standard Hanbury–Brown and Twiss setup. The sample was excited by a supercontinuum laser (SuperK EXTREME, NKT Photonics) tuned to 550nm, and the emission was detected with a pair of silicon avalanche photodiodes (τ-SPAD, PicoQuant). Detection events were recorded and correlated with a time-correlated single-photon counting (TCSPC) module (PicoHarp300, PicoQuant).

DATA AVAILABILITY

The authors confirm that the data supporting the findings of this study are available within the article and its supplementary materials.

METHODS-ONLY REFERENCES

[1] A. Gopinath, P.W.K. Rothemund, ACS nano, 8 (2014) 12030-12040.
[2] D. Nečas, P. Klapetek, Central European Journal of Physics, 10 (2012) 181-188.
[3] S.M. Douglas, A.H. Marblestone, S. Teerapittayanon, A. Vazquez, G.M. Church, W.M. Shih, Nucleic Acids Res., 37 (2009) 5001-5006.

**Supplementary Information**

**Contents**

**Supplementary Note 1.**    DNA origami "anchors" design

The 'sameside sharp triangle' design is taken from [61]. We modified 18 staples near the hole of the triangle by extending them with 20-nucleotide adenine tails at the 3′ end (see Supplementary Figure 1). These extended strands consistently point toward the upper side of the surface-bound triangle.

**Supplementary Note 2.**    Design of the DNA patterns and AFM characterization

In this study we employed two distinct patterns for DNA triangle assembly, detailed in Supplementary Figures 2-3.
• Pattern 1: A 25 μm square array of 120 nm triangular binding sites with a 170 nm period (Supplementary Figure 2).
• Pattern 2: A 20 μm × 25 μm pattern featuring square arrays of DNA triangles, with periods gradually decreasing from 1000 nm to 170 nm (Supplementary Figure 3).

Following placement, chips containing functionalized DNA triangles were dried and imaged using dry-mode atomic force microscopy (AFM; Dimension ICON AFM, Bruker). Each AFM image was processed using the Gwyddion software [85]. AFM measurements confirmed a DNA layer height of approximately 2 nm across all patterns, consistent with the expected height of a DNA double helix of 2 nm (Supplementary Figures 2-3).

**Supplementary Note 3.**    Fitting of photoluminescence of localized and free exciton

Assuming Lorentzian peak shapes, the PL intensity can be estimated as:

$$I(E, n) \propto n c_L \frac{\gamma_L}{(E - E_L)^2 + \gamma_L^2} + (1 - n) c_F \frac{\gamma_F}{(E - E_F)^2 + \gamma_F^2},$$

where $E$ is the photon energy, $E_L$ and $E_F$ are energies of the localized and free excitons, $\gamma_L = 34$ meV and $\gamma_F = 24.5$ meV are half widths at half maximum of respective peaks, $n$ is the relative density of localized and free exciton states, $c_L$ and $c_F$ are factors accounting for oscillator strengths and the thermal distribution of localized and free excitons.

To compare the model with experimental data, we calculate the PL intensities at the resonant energies:

$$I(E_L, n) \propto \frac{n c_L}{\gamma_L} + \frac{(1 - n) c_F \gamma_F}{\Delta^2 + \gamma_F^2},$$

$$I(E_F, n) \propto \frac{n c_L \gamma_L}{\Delta^2 + \gamma_L^2} + \frac{(1 - n) c_F}{\gamma_F},$$

where $\Delta = E_F - E_L$ is 50 meV. Because the relative density $n$ is small, we use the limit $n \to 0$, and estimate the ratio of intensities at the resonant energies as

$$R = \frac{I(E_L, n)}{I(E_F, n)} = \frac{\gamma_F^2}{\Delta^2 + \gamma_F^2} + \frac{\gamma_F \Delta^2 (\Delta^2 + \gamma_F^2 + \gamma_L^2)}{\gamma_L (\Delta^2 + \gamma_F^2)(\Delta + \gamma_L^2)} \frac{c_L}{c_F} n + O(n^2)$$

In experiment we know the density of DNA triangles $N$, and we assume $n = \alpha N$ with $\alpha$ is a proportionality coefficient. Substituting this into the equation, we obtain:

$$R = \frac{\gamma_F^2}{\Delta^2 + \gamma_F^2} + \frac{\gamma_F \Delta^2 (\Delta^2 + \gamma_F^2 + \gamma_L^2)}{\gamma_L (\Delta^2 + \gamma_F^2)(\Delta + \gamma_L^2)} \frac{c_L}{c_F} \alpha N,$$

For linear fitting, we define a single parameter $\beta = c_L \alpha / c_F$. Substituting experimental data, we find:

$$R \approx 0.19 + 0.68\beta N.$$

Fitting this expression gives $\beta = 0.1 \, \mu m^2$. This simple linear fit demonstrates the scalability of our approach.

Supplementary Figure 1: Design of the triangle extensions

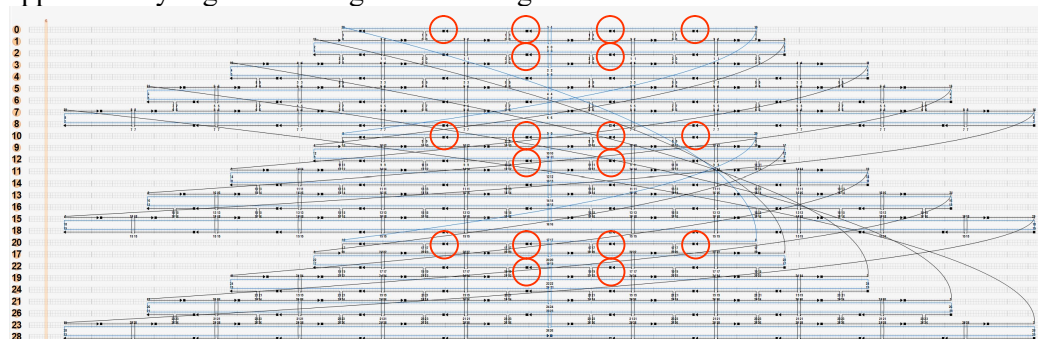

The staple diagrams of the triangle exported from cadnano 2.2.[62] Red circles indicate the 18 staples on the triangle which are extended by 20 adenine nucleotides at the 3′ end.

Supplementary Figure 2: Nanopattern 1 design and AFM characterization

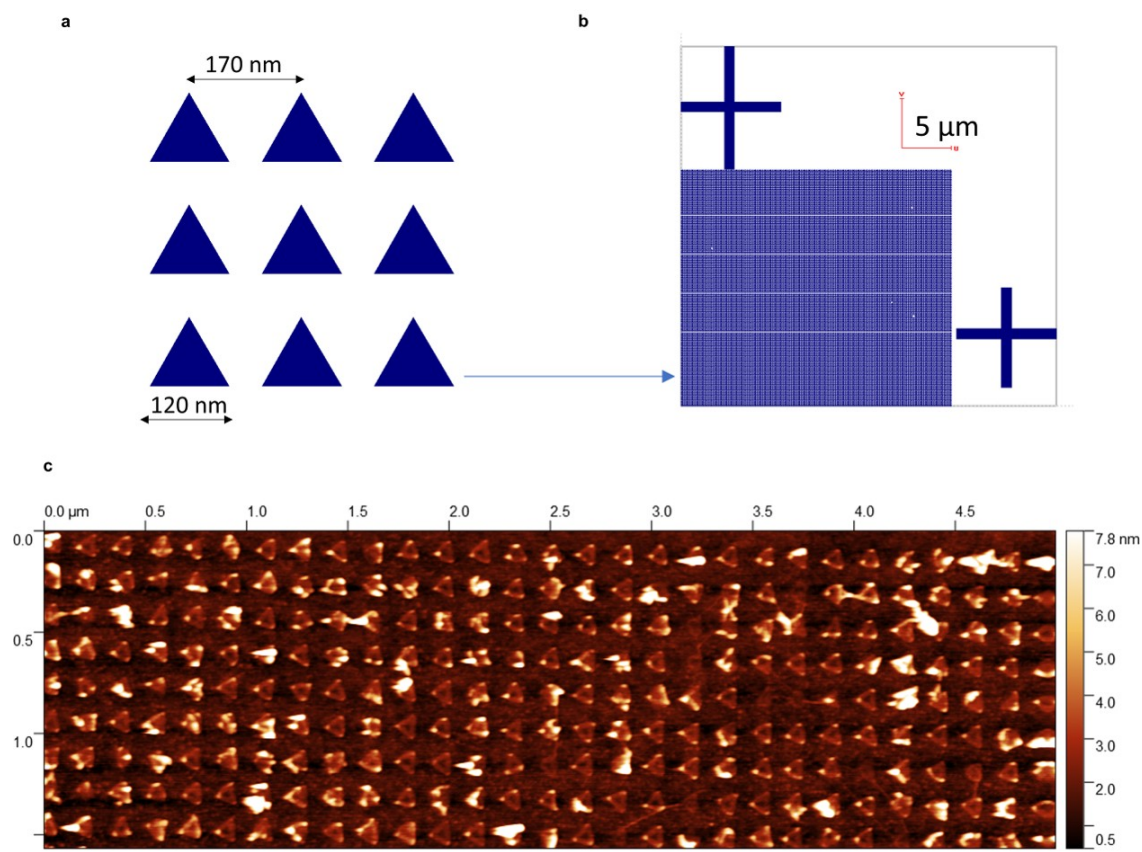

**a, b** Schematic of a patterned area. Crosses are used as fiducial markers. The nanopatterned area is 25 μm x 25 μm and consists of arrays of triangular binding sites with size of 120 nm in a square lattice with 170 nm period. Panel **a** shows a zoomed-in area of the lattice. **c** Typical dry-mode AFM images of Si/SiO$_2$ patterned chips with triangles bearing 18 T$_{19}$ extensions with 170 nm period between triangles. The AFM image size is 1.5 μm × 5 μm.

Supplementary Figure 3: Nanopattern 2 design and AFM characterization

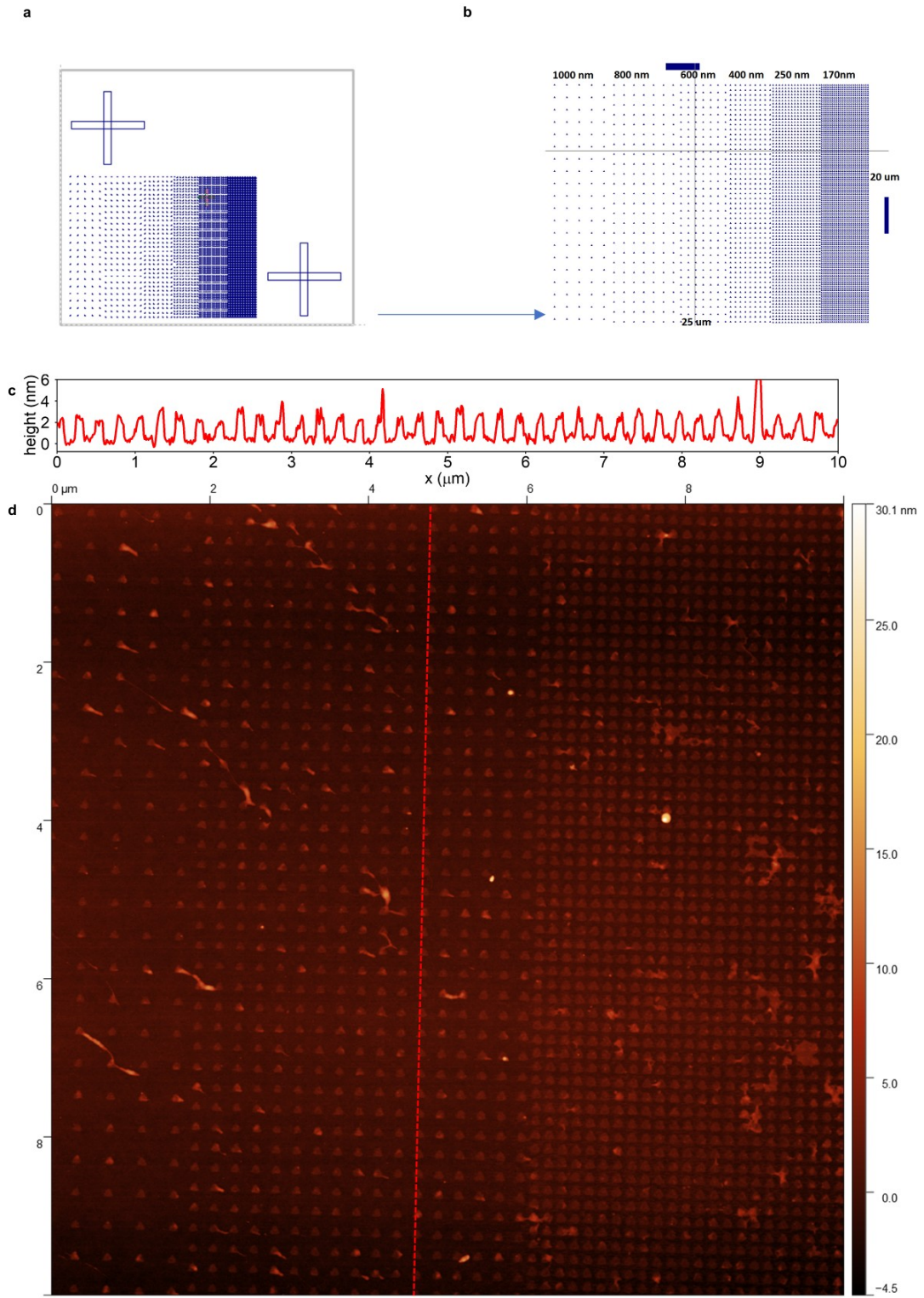

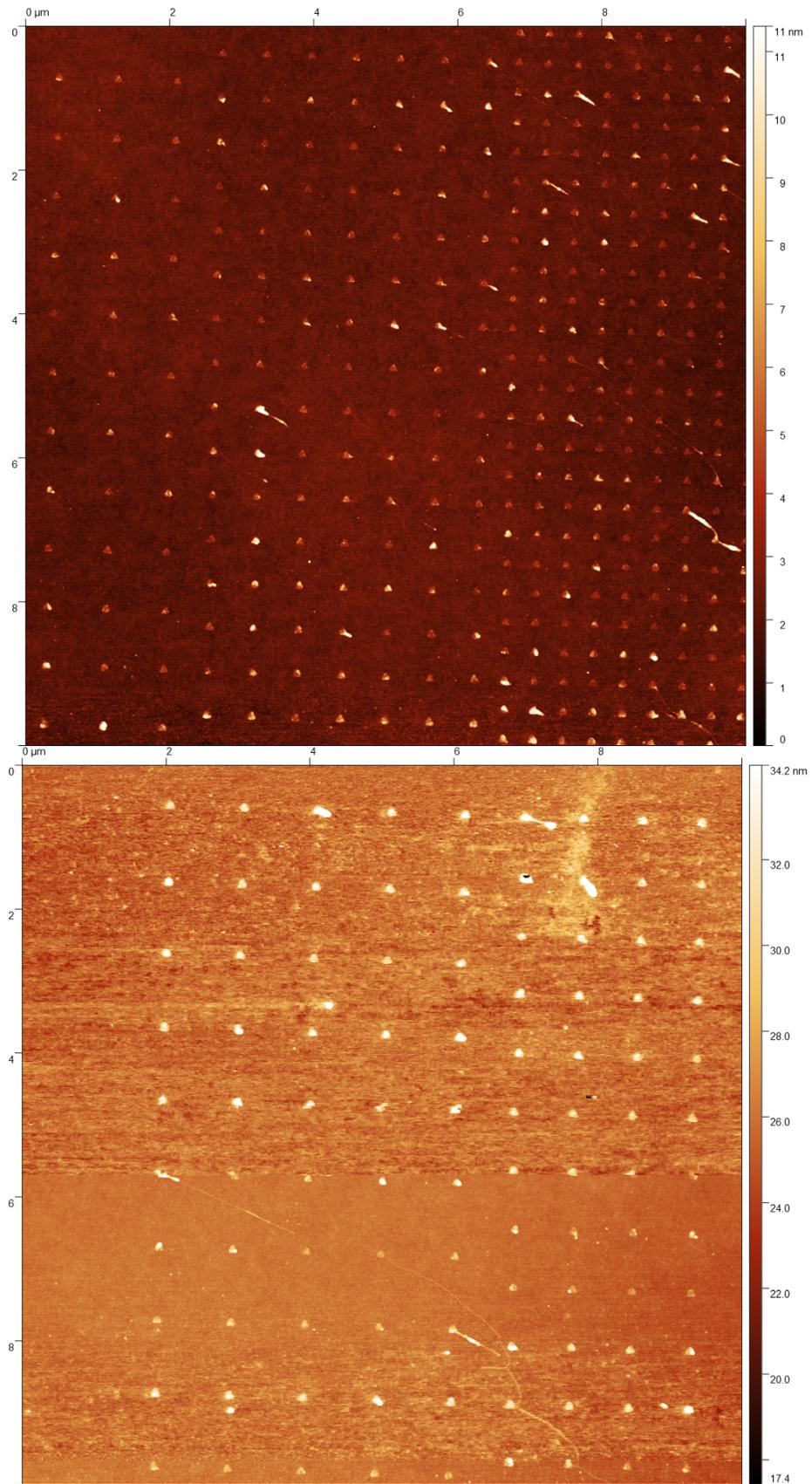

**a, b** Schematic of a patterned area. Crosses are used as fiducial markers. The nanopatterned area is 20 µm x 25 µm and consists of arrays of triangular binding sites with size of 120 nm in a square lattice with periods gradually decreasing from 100 nm to 170 nm. **c** Typical height profile from dry-mode AFM images of Si/SiO₂ patterned chips with triangles bearing 18 T₁₉ extensions with thiol group on its 5' end. **d** Typical dry-mode AFM images of Si/SiO₂ patterned chips with triangles bearing 18 T₁₉ extensions thiol group on its 5' end. The AFM images size is 10 µm × 10 µm.

Supplementary Figure 4: DNA origami-programmable quantum light sources in MoS₂

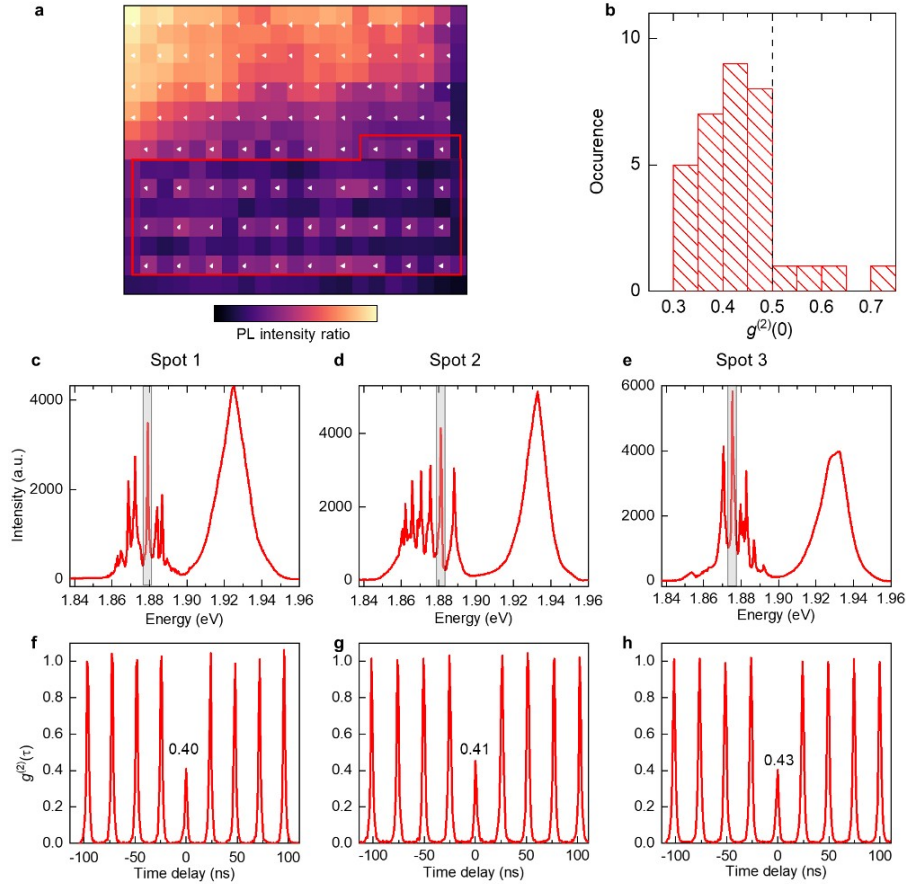

(**a**) Map of localized/free exciton PL peaks ratio, corresponding to the marked area in panel **c** in Fig. 2 in the manuscript. Each white triangle indicates the position of a DNA origami triangle. (**b**) A statistical histogram shows the second-order coherence at zero-time delay, measured at 33 spots, each corresponding to the position of a DNA origami triangle bearing 54 T19 extensions with thiol group on its 5' end. These 33 measured spots are marked by red line in panel (**a**). (**c**, **d**, **e**) Low temperature 4K PL spectra from the three measured spots. (**f**, **g**, **h**) Correlation functions of the localized defect states indicated by the gray bars in panels (**c**), (**d**), and (**e**). The values of the second order coherence at zero time delay, $g^{(2)}(0)$, are indicated in each correlation function plot.

Supplementary Figure 5: Photostability of DNA origami-programmable quantum light sources in $MoS_2$

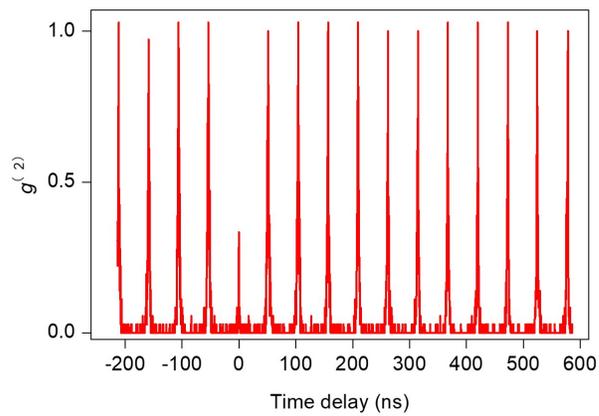

The full-timescale photon correlation function is presented in Fig. 3d of the manuscript. The absence of bunching at long delays (up to 600 ns) rules out blinking on timescales shorter than 600 ns.

Additional References

[1] A. Gopinath, P.W.K. Rothemund, ACS nano, 8 (2014) 12030-12040.

[2] D. Nečas, P. Klapetek, Central European Journal of Physics, 10 (2012) 181-188.

[3] S.M. Douglas, A.H. Marblestone, S. Teerapittayanon, A. Vazquez, G.M. Church, W.M. Shih, Nucleic Acids Res., 37 (2009) 5001-5006.